\documentclass[twocolumn,aps,prl]{revtex4}
\usepackage[french]{babel}
\usepackage[latin1]{inputenc}
\usepackage[OT1]{fontenc}
\usepackage{amsfonts}
\usepackage{epsfig}
\usepackage{amsmath}
\usepackage{amssymb}

\newcommand{\cc}{{\cal C}} 
\newcommand{\dd}{{\cal D}} 
\newcommand{\tc}{t_{cross}}

\begin{document}
\title{Aging is - almost - like equilibrium}

\author{Alexandre Lef\`evre}
\affiliation{Institut de Physique Th\'eorique,
CEA, IPhT, F-91191 Gif-sur-Yvette, France\\
CNRS, URA 2306, F-91191 Gif-sur-Yvette, France}
\email{alexandre.lefevre@cea.fr}
\date{\today}

\begin{abstract}
We study and compare equilibrium and aging dynamics on both sides of the ideal glass transition temperature $T_{MCT}$.  In the context of a mean field model, we observe that all dynamical behaviors are determined by the energy distance $\epsilon$ to threshold - i.e. marginally stable - states. We furthermore show the striking result that after eliminating age and temperature at the benefit of $\epsilon$, the scaling behaviors above and below $T_{MCT}$ are identical, reconciling {\it en passant} the mean field results with experimental observations. In the vicinity of the transition, we show that there is an exact mapping between equilibrium dynamics and aging dynamics. This leads to very natural interpretations and quantitative predictions for several remarkable features of aging dynamics: waiting time-temperature superposition, interrupted aging, dynamical heterogeneity.
\end{abstract}

\maketitle

\section*{Introduction}
Glassy systems are characterized by very slow relaxations and
non-equilibrium behavior. In general, there is however a high
temperature (or low density, pressure, etc) equilibrium phase where
all quantities relax to their equilibrium value in finite time. Between the two is the putative
glass transition, which has been identified as a true singularity
in the free energy in spin glasses or in the dynamic free energy in
mean field models of glass formers. 
The main predictions of mean field models concerning the glass phase are the existence of an effective temperature $T_{eff}$ at which the slowest degrees of freedom are ``thermalized'' and a specific form for the decay of the correlators at long times. Although effective temperatures have been reported in experiments and numerical simulations~\cite{makse,kob}, the time evolution predicted by mean field models is in general not observed. 
There is thus still a deep challenge in interpreting the measurements of slow relaxations in the glass phase as well as in describing the crossover between aging and equilibrium dynamics
which take place when one is patient enough to wait for the system to reach equilibrium. Among many interesting features of glassy materials, waiting time-temperature superposition (WTTS) has been reported in various glass formers~\cite{struick,oconnell,lunk}. 

The aim of this Letter is, in the context of a mean field model, to answer the following important questions. Are mean field predictions compatible with experimental observations ? Are slow dynamics in the glass phase intrinsically different from slow dynamics in the equilibrium phase ? How does the crossover from aging to equilibrium occur at the critical point ? What is the origin of the observed WTTS ? The answer to these questions we will shed new light on the central role played by the relaxation of the energy.

The Letter is organized as follows: we first describe the model and the old solution of its dynamics, and then improve it with some new results. Next, we establish a correspondence between slow dynamics at equilibrium close to $T_{MCT}$ and aging dynamics below $T_{MCT}$, which becomes an exact mapping at $T_{MCT}$. The consequences of this correspondence, WTTS, interrupted aging and dynamical heterogeneity, are then discussed.
Some results presented here follow from long and technical derivations, which will not be given here and will be detailed somewhere else~\cite{companion}.

\section{The mean field picture of aging}
We start discussing the dynamics of the spherical p-spin model, where exact statements can be made~\cite{cuku}. 
It can be seen as a ground for establishing rigorous results within the landscape approach following Goldstein~\cite{goldstein}, and developed by many authors - see~\cite{cavagna,sastry} for instance. The model consists of $N$ variables $S_i$ subject to the constraint $\sum_i S_i^2=N$, with Hamiltonian $H=\sum_{i_1,\cdots,i_p}J_{i_1,\cdots,i_p}S_{i_1}\cdots S_{i_p}$, where the $J$'s are quenched Gaussian random variables. Explicit equations can be written for the dynamics of correlators $C(t,t')=\langle S_i(t)S_i(t')\rangle$ and response functions $R(t,t')=\langle\delta S_i(t)/\delta h_i(t')\rangle$ to an external field. The solution of these equations, which can be found in~\cite{cuku}, captures the features of glassy dynamics, where two well separated times scale emerge, corresponding to different kinds of degrees of freedom, some - ''local'' - responding as in equilibrium, and others - ''structural'' - being out of equilibrium.
In the equilibrium time sector, fluctuation dissipation theorem (FDT) holds. In the long time aging sector, FDT does not hold, but instead a modified form of it, where the bath temperature $T$ has to be replaced by an effective temperature $T_{eff}>T$ which can be computed explicitly. 
A fine analysis shows that slow degrees of
freedom are indeed thermalized at $T_{eff}$ rather than $T$, coming from the contribution of the structural degrees of freedom to the entropy at some energy $E_{Th}$. The threshold energy $E_{Th}$ is the energy separating unstable and stable states, and corresponds to the asymptotic energy: at long times, a glassy system is close to marginal stability.  
In the equilibrium phase, the energy is $E_{eq}(T)$ is above $E_{Th}(T_{MCT})$, and at the approach of $T_{MCT}$, the system gets closer to the threshold and slows down, almost trapped by marginally stable states. Similarly, in the glass phase, where the equilibrium energy is below the threshold, the system loses equilibrium when it gets close to $E_{Th}(T)$ and thus ages. In both cases, the closer the system is from the threshold level, the slower it is~\cite{laloux}. Before making this more precise, we briefly recall the known scalings for the correlator $C(t,t_w)$  and the response function $R(t,t_w)$, using as usual $\tau=t-t_w$, in the equilibrium and glass phases~\cite{cuku}. 

\subsection{The old results}
In the equilibrium phase, i.e. $T=T_{MCT}+\epsilon$,
there are two characteristic time scales, the $\beta$-decay around the plateau and the $\alpha$-relaxation~\cite{goetze}: $\tau_\beta(\epsilon)=\epsilon^{-1/2a}$ and $\tau_\alpha(\epsilon)=\epsilon^{-\gamma}$, $\gamma=1/2a+1/2b$, where $a$ and $b$ are well known MCT exponents. There, the scalings for $C$ are, in order: $C(t,t_w)=q_c+\sqrt{\epsilon}c_\beta(\tau/\tau_\beta)$ and $C(t,t_w)=q_cc_\alpha(\tau/\tau_\alpha)$, where $q_c$ is the non-ergodicity parameter at $T_{MCT}$. 

In the glass phase $T<T_{MCT}$, two characteristic time sectors have also been identified~\cite{cuku}. The first time sector is stationary: when $\tau\ll t_w$, there is time translation invariance (TTI): $C(t,t_w)=C_{ST}(\tau)$ and FDT is still valid: $R(t,t_w)=\frac{1}{T}\frac{\partial C(t,t_w)}{\partial t_w}$. The second - aging - time sector corresponds to $\tau\sim t_w$~\cite{incorrect}. There, FDT must be modified: $R(t,t_w)=\frac{1}{T_{eff}}\frac{\partial C(t,t_w)}{\partial t_w}$ and the scaling is $C(t,t_w)=\cc(h(t_w)/h(t))$. An interesting but unpleasant feature is that the analysis of the aging equations at infinite $t_w$ is not sufficient to determine $h(t)$~\cite{cuku}, which has remained unknown until recently.
From this point of view, equilibrium and aging dynamics seem to differ strongly. We now improve an analysis made recently of the dynamical equations at large but finite $t_w$~\cite{andreanov} and show that in fact both phases are very similar.

\subsection{New results about the aging regime}
Recently, it was shown that for $T<T_{MCT}$, aging actually sets in around the plateau~\cite{kim,andreanov}, a time sector which is usually just considered as the matching point of the TTI and  aging sectors. 
It was shown that in this regime, the correlators have a scaling form $C(t,t_w)=q+t_w^{-\alpha}g(\tau/t_w^\beta)$, where the exponents are related through $\alpha=\beta a$. In addition, matching with the aging regime, the uncertainty on the function $h(t)$ was considerably reduced to $h(t)=\exp\left(A\frac{t^{1-\mu}}{1-\mu}\right)$, where $A$ is a constant which can be safely absorbed into the relaxation time and set to $1$, while the aging exponent $\mu\leq 1$ verifies $\mu b=\alpha+b \beta$ and now $a$ and $b$ are aging MCT exponents. Remark that $\mu\leq 1$ is required for this solution to be consistent. Situations where this occurs have been reported in various experiments, e.g.~\cite{viasnoff}. 
Values of $\mu$ exceeding $1$ (super-aging, see e.g.~\cite{ronsin}) are the signature of a different relaxation mechanism.
After this analysis, the complete solution is still to be found, as the exponents are unknown although all determined by $\alpha$. However, interestingly, in the same time the energy was shown to relax as $E(t)=E_{Th}+E_2 t^{-2\alpha}$, meaning that the whole slowness of the correlators is actually encoded 
in the energy relaxation towards the threshold level. This is remarkable and will lead to the striking correspondence between equilibrium and aging dynamics, which we shall establish below. Last, but not least, the scaling of the response function is $R(t,t_w)=-t_w^{-\alpha-\beta}w'(\tau/t_w^\beta)/T$ and the ratio $X(y)=w(y)/g(y)$ interpolates smoothly between $X(-\infty)=1$ and $X(\infty)=X=T/T_{eff}$, showing that all interesting features of the aging regime already occur in the plateau regime, which corresponds to the time scales where motion propagates from local to structural degrees of freedom.

\section{Aging is like equilibrium}
We now analyze further and improve these results, in order to obtain a unified picture of equilibrium and glass phases.
First, we recall the simplest of the two equations verified by the scaling functions $g$ and $w$ around the plateau~\cite{andreanov} (where $x_0$ is such that $g(x_0)=0$):
\begin{equation}
\label{eq:g}
\begin{split}
\psi(x)+\int_{x_0}^xdy\, \frac{w'(y) g(y)}{T_{eff}}=\psi(x_0), \\
\psi(x)=w(x)^2+\int_0^x dy\,w'(y)\left(w(x-y)-w(x)\right)
\end{split}
\end{equation}
When $T=T_{MCT}$, $T=T_{eff}$, and Eq.~(\ref{eq:g}) reduces to the equation for the scaling function $c_\beta$ of equilibrium dynamics~\cite{goetze}. Second, 
writing $h(t)=\exp\phi(t)$, it is easy to verify that in order to have $\cc(h(t_w)/h(t))$ be of order $1$, one needs that $\tau\sim \phi'(t_w)=t_w^\mu$. This has an important consequence: the scaling $\cc(h(t_w)/h(t))$ with a stretched exponential $h(t)$ is in general very hard to observe except in the p-spin model~\cite{kim}, and in the large $t_w$ limit it differs only through tiny asymptotic corrections to the observed scaling $C(t,t_w)=\tilde{\cc}(\tau/t_w^\mu)$. This explains the many reported discrepancies between mean field predictions and measurements~\cite{makse,kob}. Third, and more strikingly, the aging relaxation can also be related to the scaling function of the $\alpha$-regime in the equilibrium phase, {\it at all temperatures}. Indeed, the scaling function $\cc$ was found in~\cite{cuku} to verify
\begin{equation}
  \label{eq:cag}
\begin{split}
 qX \int_\lambda^1d\lambda'\,\frac{d }{d\lambda'}\cc(\lambda')^{p-1}\cc\left(\frac{\lambda}{ \lambda'}\right)=\\
(p-1)(1-q)\cc(\lambda)-\cc(\lambda)^{p-1}(1-q+qX).
\end{split}
\end{equation}
We now show that the solution of this equation is in fact a familiar function of the equilibrium phase. Writing $\lambda=\exp(-\tau)$, one gets, after a bit of algebra: 
\begin{equation}
  \label{eq:ag2}
  \dd(t)+\frac{pq_c^{p-1}}{2T_{MCT}^2}\int_0^\tau dt \,\dd^{p-1}(t)\dd'(\tau-t)=0,
\end{equation}
with $\dd(x)=\cc(\exp(-x))$, which is exactly the equation verified by $c_\alpha$~\cite{goetze,remark}. Using $h(t)\approx\exp\left(\tau/t_w^\mu\right)$, and remarking that the small $\tau$ behavior must match the $\beta$-regime, we can determine unambigously $\dd$, leading to: $C(t,t_w)=q c_\alpha\left(\tau/\tau_\alpha(t_w)\right)$, where $\tau_\alpha(t_w)\propto t_w^\mu$.


\subsection{Mapping equilibrium onto aging}
The interpretation of the preceding paragraph will be central result. We introduce $\epsilon(T,t_w)=E(t_w)-E_{Th}(T)$
and obtain for $T<T_{MCT}$: $\tau_\beta(t_w)=\epsilon(T,t_w)^{-1/2a}$ and $\tau_\alpha(t_w)=\epsilon(T,t_w)^{-1/2a-1/2b}$. Added to the above scalings, this leads us to the conclusion that {\it once temperature and $t_w$ dependences are expressed in terms of $\epsilon(T,t_w)$ only, equilibrium and aging dynamics almost map onto each other.} Almost, because $a$, $b$ and the scaling functions in the $\beta$-regime explicitly depend on $T$ below $T_{MCT}$. However, the mapping becomes exact at the transition, and is very accurate close to it (see Fig.~\ref{fig}). This is a rather unexpected and spectacular result, as it gives a route to studying aging dynamics from the knowledge of the deeply supercooled equilibrium phase - calculations in both phases are not intrinsically different. In addition, it incorporates in a natural way the WTTS reported in experiments.
\subsection{Interrupted aging}
An immediate consequence of that is interrupted aging. Indeed, the glass transition temperature $T_g$ is where a glass former seems to lose equilibrium, but where, with a bit of patience, it is still possible to have it equilibrate. More precisely, there is a crossover waiting time $\tc$ such that for $t_w<\tc$ aging occurs, while for $t_w>\tc$, the system has reached equilibrium and its dynamics is TTI. By definition of $T_g$, $\tc$ is large. A na\"\i ve guess for the value of $\tc$ would be $\tc=\tau_{\alpha_c}$, the equilibrium relaxation time. However, this is too na\"\i ve, as this is the time for the system to have visited several equilibrium states, while $\tc$ is the time where aging stops and is in principe much smaller. In our model, it is possible to compute $\tc$ using the previous analysis, from the following gedanken experiment: quench the system from $T=2T_{MCT}$ down to $T=T_{MCT}+\varepsilon$, $\varepsilon\ll T$. During the first stage of the dynamics, the energy is close but not enough to its equilibrium value and thus relaxes as a power law: $E(t_w)-E_{Th}(T_d)\approx t_w^{-2\alpha_c}$ and the system ages. This ends when the energy reaches its equilibrium value $E(T)$, which gives: $\tc\sim |T-T_{MCT}|^{-1/2\alpha_c}$.
This leads to the following scaling form for the $\alpha$-relaxation time: $\tau_\alpha=\tau_{\alpha_c}^{Eq}(T){\cal T}(t_w/\tc)$,
where ${\cal T}(x)\sim  x^{\mu_c}$ at $x\ll 1$ and ${\cal T}(\infty)=1$. Remark that $\tc$ may also be expressed as $|T-T_{MCT}|^{\gamma/\mu_c}$, that is the exponent is the ratio of the ones of the relaxation time in both phases, which gives an easier way to measure it in the MCT regime.

\subsection{Dynamical heterogeneity}
\begin{figure}
\includegraphics[width=.4\textwidth]{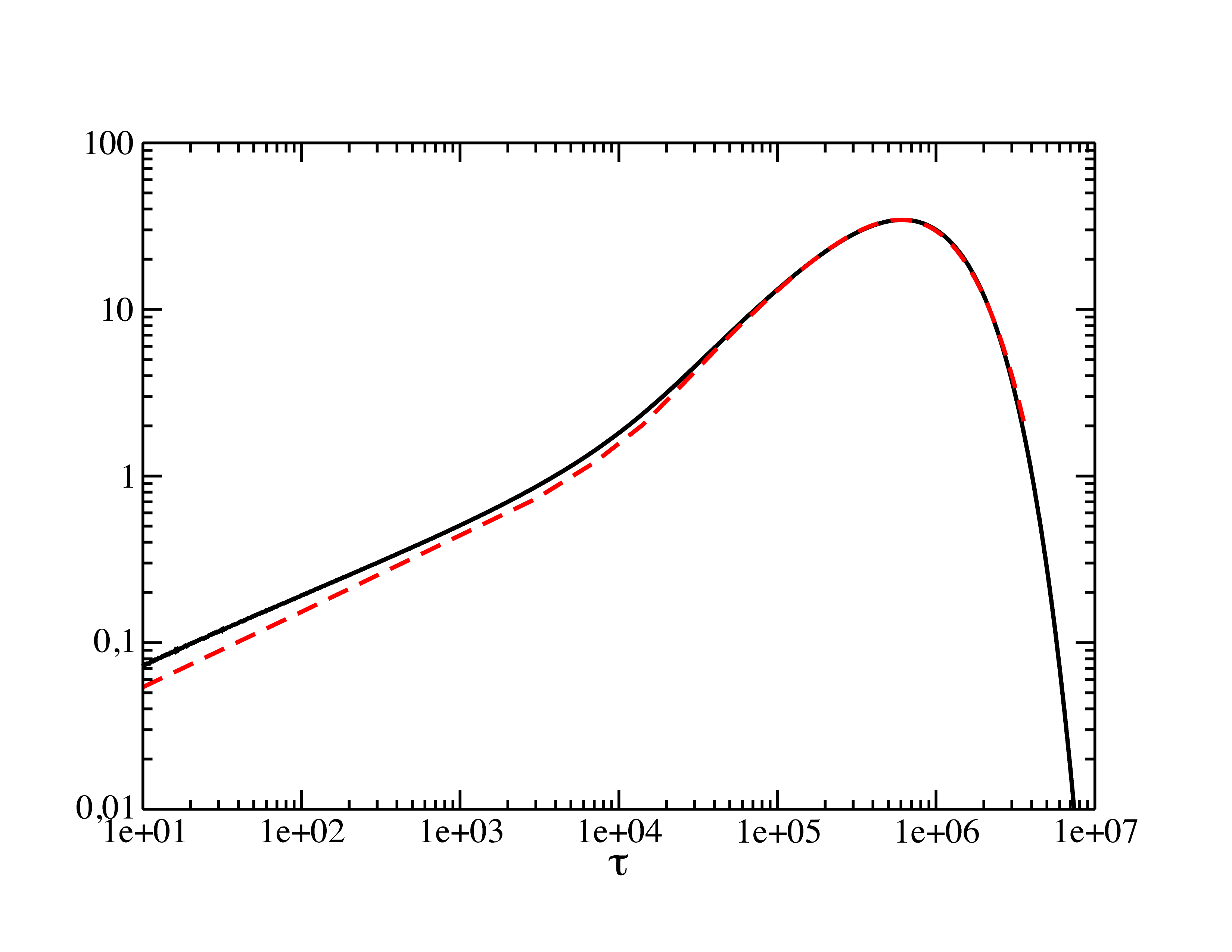}
\caption{\label{fig} Comparison of $\chi_T(\tau)$ above $T_{MCT}$ (straight line) and $\chi_w(\tau)$ well below $T_{MCT}$ (dashed line), with the same value of $\epsilon$. The Curves have been rescaled vertically, as the factor $\partial \epsilon(T,t_w)/\partial t_w$ occurring in $\chi_w$ is not known.}
\end{figure}

Using the previous results, it is possible to describe the general behavior of correlation functions generally used for describing dynamical heterogeneities (DH). We start with 
the now widely studied $\chi_4(t,t_w)=\int dy \left(\langle \delta C(x,t,t_w)\delta C(x+y,t,t_w)\rangle-{\langle \delta C(x,t,t_w)\rangle}^2\right)$, where $\delta C(x,t,t_w)$ is a local correlator at position $x$. It is possible~\cite{companion} to extend the field-theoretic approach to $\chi_4$ used by Biroli and Bouchaud~\cite{biroli} to the aging regime. It results that $\chi_4(t,t_w)$ is a sum of ladder diagrams built with the full correlator $\delta C(x,t,t_w)$, which time dependence follows that of the previous paragraphs. This has several immediate consequences. First, all scaling results given in previous analysis of $\chi_4$ in the equilibrium phase can be directly applied to the glass phase, provided we replace $\epsilon$ by $\epsilon(T,t_w)$, providing MCT predictions for DH in he glassy phase. Second, it was reported in~\cite{castillo} that, once $\chi_4(t,t_w)/\underset{\tau}{\mathrm{max}}\chi_4(t,t_w)$ is plotted versus $1-C(t,t_w)$, all data - i.e. for all $t_w$'s - collapse onto a single master curve. This could be obtained from general scaling considerations, but interestingly, it comes out naturally from the fact that all $t_w$-dependance of both quantities plotted is through $\epsilon(T,t_w)$, which gives the natural parameterization of the master curve. In addition, close to $T_{MCT}$, the full master curve should co\"\i ncide with the one of the equilibrium phase.
Another quantity which has been shown to be interesting above $T_{MCT}$ is $\chi_T(\tau)=\frac{\partial C(\tau)}{\partial T}$, which has been shown to have the same critical behavior as $\chi_4(\tau)$, and which is more easily accessed experimentally~\cite{chiT}. It is very simple to generalize this quantity to the glass phase, as $\partial/\partial T$ is identical to $\partial/\partial E$, up to the non-singular multiplying factor $\partial E/\partial T$. It is thus natural to generalize $\chi_T$ to:
\begin{equation}
  \label{eq:chiw}
  \chi_w(\tau+t_w,t_w)=\frac{\partial C(\tau+t_w,t_w)}{\partial \epsilon(T,t_w)}.
\end{equation}
In the equilibrium phase, $\chi_w(t,t_w)$ reduces to $\chi_T(\tau)$, while in the glass phase, it becomes $\left(\frac{\partial \epsilon(T,t_w)}{\partial t_w}\right)^{-1}\frac{\partial C(\tau+t_w,t_w)}{\partial t_w}$. This expression makes $\chi_w$ as easy to measure in experiments as $\chi_T$. In Fig.~\ref{fig}, $\chi_w$ at $T=T_{MCT}+10^{-3}$ and $\chi_T$ at $T=.5\approx T_{MCT}-0.125$, with approximately the same value of $\epsilon$ are plotted versus $\tau$. Clearly, the peaks of the $\alpha$-relaxation co\"\i ncide very well, while the $\beta$-regime slightly differ, having two different exponents, respectively $a\approx 0.396$ and $a\approx 0.448$. When the same procedure is made at $T=T_{MCT}$ for $\chi_w$, the curves fall on top of each other.

\section{Conclusion}
In this Letter, we have shown that, within a mean field model, dynamics in the ideal glass phase are essentially identical to equilibrium dynamics just above the glass transition, and that all important features in both phases may be absorbed in a single function $\epsilon(T,t_w)$, with important physical significance, being the distance in energy to threshold states.
This is a remarkable result, as it opens the door to interpreting the aging data obtained in the glass phase of molecular or collo\"\i dal glass formers. In particular, it provides general scaling laws for multipoint correlation functions - relaxation, dynamical heterogeneities - as well as the natural parameterization to seek, $\epsilon(T,t_w)$, which may alternatively be an enthalpy difference. It also naturally predicts WTTS, which was reported in aging measurements~\cite{lunk}. On the conceptual aspect, it is very unexpected, because it shows that aging dynamics is actually very similar from equilibrium dynamics in the deeply supercooled regime, which is rather far from common thinking about aging. But this is not so surprising, and reflects the fact that both in liquid and glass phases, slow dynamics occurs because of the roughness of the energy landscape, which shape does not change qualitatively more than static quantities when crossing the glass transition.
Here, several strong predictions have been made, which should be tested in experiments and numerical simulations. Doing so, one must keep in mind that in general, the exponents $a$ and $b$ both depend on temperature, and thus the predictions made here should in general be tested at close temperatures.
This would also be of valuable help for determining situations where the energy landscape approach gives a qualitatively correct picture. On this prospect, and more speculatively, one may ask whether the effect of activation at low temperature may be also reabsorbed in the same way. It would be also worth coming back to data from previous experiments, where disagreement with former mean field predictions where found, and reinterpret them using the predictions given in this Letter. 

\section*{Aknowledgements}
This work is supported by the ANR grant DynHet. The author thanks G. Biroli and J.-P. Bouchaud for stimulating discussions and careful reading of the manuscript, as well as J. Kurchan. This article was written during a stay at the Columbia University, NY, and D. Reichman is warmly thanked for his hospitality. The numerical checks have been made using a code provided by K. Miyazaki.


\begin{thebibliography}{99}
\bibitem{remark} This was also noticed in~\cite{ledou}, but not the exact connexion to equilibrium dynamics.
\bibitem{incorrect} This is actualy not correct when $\mu<1$, as shown later in the text, but we recall it from Ref.~\cite{cuku}.
\end{thebibliography}
\end{document}